# BATMAN—an R package for the automated quantification of metabolites from NMR spectra using a Bayesian Model

Jie Hao[1], William Astle[2], Maria De Iorio[3], and Timothy Ebbels[1,*]

[1] Biomolecular Medicine, Department of Surgery and Cancer, Faculty of Medicine, Sir Alexander Fleming Building, Imperial College London, London, SW7 2AZ, UK.

[2] Department of Epidemiology, Biostatistics and Occupational Health, McGill University, Purvis Hall, 1020 Ave. des Pins Ouest, Montreal, QC, H3A 1A2, Canada.

[3] Department of Statistical Science, University College, Gower Street, London, WC1E 6BT, UK.

**ABSTRACT**
**Motivation:** NMR spectra are widely used in metabolomics to obtain metabolite profiles in complex biological mixtures. Common methods used to assign and estimate concentrations of metabolites involve either an expert manual peak fitting or extra pre-processing steps, such as peak alignment and binning. Peak fitting is very time consuming and is subject to human error. Conversely, alignment and binning can introduce artefacts and limit immediate biological interpretation of models.
**Results:** We present the Bayesian AuTomated Metabolite Analyser for NMR spectra (*BATMAN*), an R package which deconvolutes peaks from 1-dimensional NMR spectra, automatically assigns them to specific metabolites from a target list and obtains concentration estimates. The Bayesian model incorporates information on characteristic peak patterns of metabolites and is able to account for shifts in the position of peaks commonly seen in NMR spectra of biological samples. It applies a Markov Chain Monte Carlo (MCMC) algorithm to sample from a joint posterior distribution of the model parameters and obtains concentration estimates with reduced error compared with conventional numerical integration and comparable to manual deconvolution by experienced spectroscopists.
Availability: http://www1.imperial.ac.uk/medicine/people/t.ebbels/
**Contact:** t.ebbels@imperial.ac.uk

## 1 INTRODUCTION

NMR spectroscopy is widely used in metabolomics to provide information on metabolite profiles of complex biological mixtures, such as biofluids or tissue samples. These spectra commonly contain around a thousand peaks from possibly hundreds of metabolites. The concentrations of metabolites are useful in biomarker discovery, investigating the aetiology of disease and in understanding fundamental biological processes. Quantitative measurement of metabolite abundances from NMR spectra is complicated due to shifting peak positions, peak overlap, noise and effects of the biological matrix.

Excellent solutions to manual peak fitting exist (Weljie, et al., 2006); however, this option is time consuming and is subject to human error and bias (Tredwell, et al., 2011). The R package BQuant (Zheng, et al., 2011) attempts to quantify metabolites from NMR spectra using Bayesian model selection. However, it is a two step procedure which requires spectra to be pre-aligned and partitioned into bins. Consequently, it is awkward to apply as it is not fully automated. Peak shifts are not incorporated into the statistical model, which may lead to suboptimal estimates of metabolite concentrations, especially in crowded regions.

Here we present the R package *BATMAN*, which implements a Bayesian model for $^1$H NMR spectra (Astle, et al., 2012) and a Markov Chain Monte Carlo (MCMC) algorithm to automate metabolite quantification from NMR spectral data. *BATMAN* is able to automatically assign peaks to metabolites given a target metabolite list and chemical shift region(s) specified by the user.

## 2 METHODS

An NMR spectrum, **y**, can be considered as a linear combination of metabolite peaks plus noise. Metabolite peaks can be labelled catalogued or uncatalogued, according to whether their characteristic peak patterns are known to the user. We propose a two component joint model for the catalogued $\mathbf{y}^c$ and uncatalogued $\mathbf{y}^u$ metabolites as follows:

$$\mathbf{y} = \mathbf{y}^c + \mathbf{y}^u + \varepsilon, \quad \varepsilon \sim N(0, \mathbf{I}/\lambda) \quad (1)$$

where **I** is the identity matrix and $\lambda$ is a scalar precision parameter.

The signal from the catalogued metabolites, $\mathbf{y}^c$, can be modelled as a weighted summation of $M$ metabolite templates,

$$\mathbf{y}^c = \sum_{m=1}^{M} \sum_{u} \mathbf{t}_{mu}(\sigma_{mu})\beta_m \quad (2)$$

where vector $\mathbf{t}_{mu}$ is a template spectrum of the $u^{th}$ multiplet belonging to the $m^{th}$ metabolite, $\sigma_{mu}$ is the chemical shift parameter for the corresponding multiplet, and $\beta_m$ is proportional to the concentration of the $m^{th}$ metabolite. The priors for $\sigma_{mu}$ and $\beta_m$ are truncated normal distributions.

The template spectrum, $\mathbf{t}_{mu}$, in turn is generated as a weighted summation of Lorentzian peaks. By default BATMAN obtains this prior information on the peak pattern from the Human Metabolome Database (HMDB) (Wishart, et al., 2009).

---
[*]To whom correspondence should be addressed.

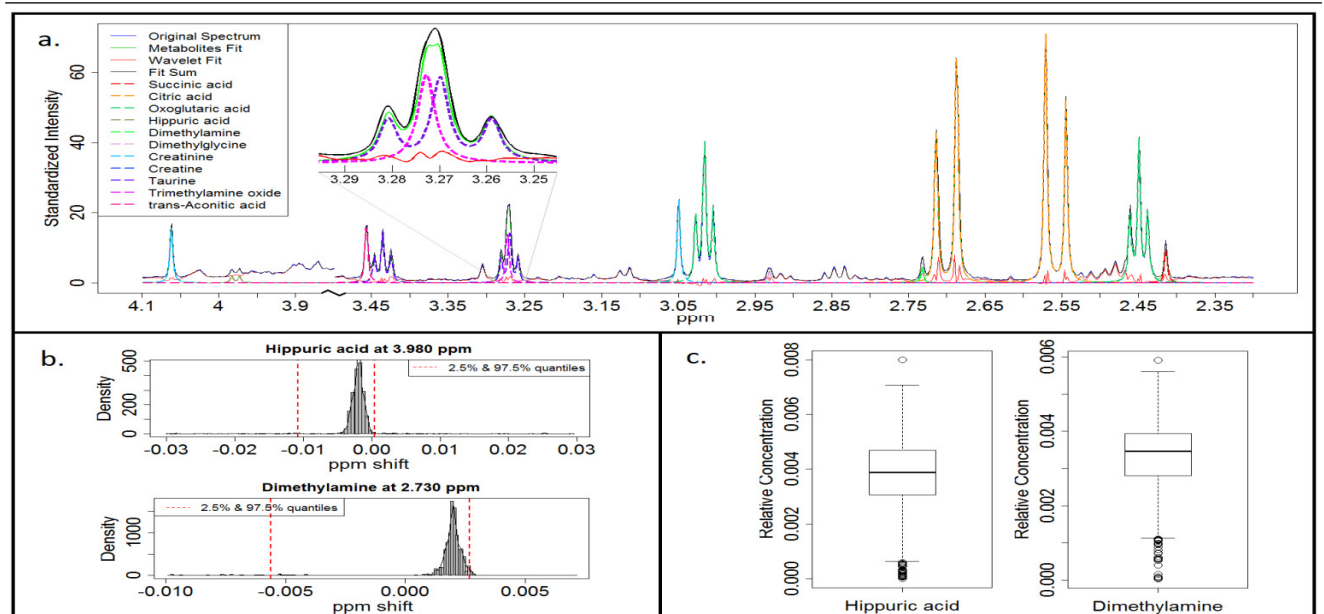

Figure 1: Typical BATMAN result on a $^1$H NMR spectrum of rat urine. a. Mean posterior fit (inset: zoom for 3.25-3.29ppm). b. Posterior distribution of chemical shift difference from catalogued position for multiplets of hippuric acid and dimethylamine. c. Posterior distribution of relative concentration for the two metabolites.

We model the Lorentzian peak width $\gamma_m$ as,

$$\ln(\gamma_m) = \mu + \nu_m \quad (3)$$

where $\mu$ represents the spectrum wide average and $\nu_m$ is a random effect, representing metabolite specific deviations from $\mu$. Our priors for $\mu$ and $\nu_m$ are both Gaussian.

Signal generated by uncatalogued metabolites, $\mathbf{y}^u$, is modelled as a linear combination of wavelet basis functions ($\mathbf{W}$). This alleviates difficulties associated with the incompleteness of the spectral library that can influence other methods (Mercier, et al., 2011). A probability density model is specified in the wavelet domain of the data, i.e. $\theta \equiv \mathbf{W} \, \mathbf{y}^u$. We use symlet-6 wavelet basis functions as they are well suited to modelling Lorentzian peaks (Astle, et al., 2012).

The distribution of $\theta$ is modelled by a truncated Gaussian, whose parameters are specified by the parameters ($\lambda$, $\psi$, $\tau$). The hyper-parameter vector $\psi \sim Gamma(c,d/2)$ allows the prior precision associated with each wavelet deviate from the global precision $\lambda \sim Gamma(a,b/2)$. $\tau$ is a truncation limit vector where each $\tau_i$ has a Gaussian distribution left truncated at a small negative value $h$.

## 3 IMPLEMENTATION

A Markov Chain Monte Carlo algorithm is proposed to sample from a joint posterior distribution of the model parameters. We temper the likelihood and penalize the wavelet component of the model stringently to move the chain into a region of good posterior support in the burn in stage of the MCMC. We introduce block updates for $\beta$ jointly with $\theta$, and for $\sigma_{mu}$ jointly with $\theta$ to allow the chain to make global moves. The details are described in (Astle, et al., 2012). Adaptation techniques are used to update the peak shift $\sigma_{mu}$ and width parameters $\gamma_m$. The core algorithm is implemented in C++ language with support for parallel processing between spectra to improve the processing speed. R version 2.12.1 (or higher), together with the R packages described in documentation, is required. BATMAN runs on Windows, Mac OSX and Linux/Unix operating systems and supports text file, R data format and 1D Bruker spectral data files as input. Processing 8 spectra on an Intel 3.0GHz Quad-Core 2 processor machine with 11 catalogued metabolites region from 2.3 to 4.1ppm took approximately 22 minutes. The MCMC procedure ran for 2000 iterations following a 4000 iteration burn in. The time required scales linearly with the number both of metabolites and spectra (on a multicore machine, multiple spectra can be analysed in parallel with the same run time as a single spectrum analysis).

## 4 RESULTS

Figure 1 shows an example result of a BATMAN fit of a $^1$H NMR spectrum of normal rat urine. The catalogued metabolite peaks are correctly fit by the algorithm, with uncatalogued peaks absorbed by the wavelet component. The unimodality of peak position and concentration distributions indicate no ambiguity in the assignments. The BATMAN deconvolution has been shown to have reduced mean estimation error compared with conventional numerical integration methods and its results on an example data set fitting 26 metabolites are comparable with the manual deconvolution by five experienced NMR spectroscopists as described by (Astle, et al., 2012).

## ACKNOWLEDGEMENTS

This work was funded by grant BB/E20372/1 from the UK Biotechnology and Biological Sciences Research Council (BBSRC).